\documentclass[%
 superscriptaddress,
 reprint,
 showpacs,
 showkeys,
 amsmath,amssymb,
 aps,
 prl,
 floatfix
]{revtex4-1}

\usepackage{graphicx}          
\usepackage{dcolumn}           
\usepackage{bm}                
\usepackage{color}
\usepackage{hyperref}          

\begin{document}
\everymath{\displaystyle}

\title{Realization of the Najafi-Golestanian microswimmer}

\author{Galien Grosjean}%
 \email{ga.grosjean@ulg.ac.be}
\affiliation{%
GRASP Lab, CESAM Research Unit, University of Li\`ege, B-4000 Li\`ege, Belgium.
}%
\author{Maxime Hubert}
\affiliation{%
GRASP Lab, CESAM Research Unit, University of Li\`ege, B-4000 Li\`ege, Belgium.
}%
\author{Guillaume Lagubeau}
\affiliation{Departamento de F\`isica, Universidad de Santiago de Chile, Chile.
}%
\author{Nicolas Vandewalle}
\affiliation{%
GRASP Lab, CESAM Research Unit, University of Li\`ege, B-4000 Li\`ege, Belgium.
}%

\date{\today}
             
\begin{abstract}
A paradigmatic microswimmer is the three-linked-spheres model, which follows a minimalist approach for propulsion by shape-shifting.
As such, it has been the subject of numerous analytical and numerical studies.
In this letter, the first experimental three-linked-spheres swimmer is created by self-assembling ferromagnetic particles at an air-water interface.
It is powered by a uniform oscillating magnetic field.
A model, using two harmonic oscillators, reproduces the experimental findings.
Because the model remains general, the same approach could be used to design a variety of efficient microswimmers.
\end{abstract}

\maketitle


The development of artificial microswimmers, microscopic robots that swim in a fluid like sperm cells and motile bacteria, could cause a leap forward in various fields such as microfluidics, microsystems or minimally invasive medicine.
At small scale, the viscous dissipation in a fluid prevails over inertia, therefore governing the swimming dynamics of microorganisms and micro-objects.
In this regime, flows are described by the Stokes equations which are linear, independent of time and therefore time-reversible.
However, in order to propel itself, \emph{i.e.} to sustain a movement without the help of an external net force, a body must produce a net flow in the direction opposite to its motion.
A microswimmer must therefore break the time-reversibility of the flow, for example by undergoing a non-palindromic sequence of deformations~\cite{lauga2009, purcell1977}.

Nature provides plenty of examples of efficient microswimmers.
For instance, microbes use their flagella, cilia~\cite{lauga2009} or the deformations of their membrane~\cite{farutin2013} to propel themselves.
Several strategies of propulsion have been studied experimentally, such as externally actuated flagella~\cite{dreyfus2005}, rotating helical tails~\cite{zhang2009} or propulsion by chemical gradients~\cite{howse2007}.
However, a bottom-up approach, looking at the simplest ingredients needed to generate a microswimmer, can lead to a deeper understanding of the swimming problem.
This approach could also provide us with designs more suited to technological or medical applications.

Simple kinematic models based on the idea of non-reciprocal deformations have been extensively studied.
The most well-known are Purcell's three-link model~\cite{purcell1977}, where three arms are linked by two hinges around which they can rotate, and Najafi and Golestanian's three-linked-spheres model~\cite{najafi2004}, where three in-line spheres are linked by two arms of varying length.
The former resembles a discrete, simplified flagellum~\cite{tam2007}. 
The latter moves by shifting mass forward, mimicking ameboids~\cite{farutin2013} and recoil swimmers~\cite{childress2011}.
It has the added advantage of involving translational degrees of freedom in one dimension, which allows analytical studies~\cite{golestanian2008}.

While these kinematic models offer a convenient basis for theoretical studies, small-scale experimental implementations are scarce.
Indeed, such models impose the shape of the swimmer at all times, as if controlled by micromotors and actuators, which leads to serious technological limitations.
In the case of the three-linked-spheres model, Leoni \emph{et al.} reproduced the deformation sequence using optical tweezers~\cite{leoni2009}.
However, this has the consequence of pinning the swimmer in a potential well, meaning that a continued translational motion is impossible~\cite{leoni2009}.

By contrast, this letter examines an experimental realization of the three-linked-spheres swimmer by self-assembly, using submillimetric spheres actuated by an external force.
Instead of rigid, extensible arms, this model is based on harmonic oscillators~\cite{lagubeau2016}.
The breaking of time-reversibility comes from a spontaneous phase shift between the oscillators.

\begin{figure}
\includegraphics[width=8.6cm]{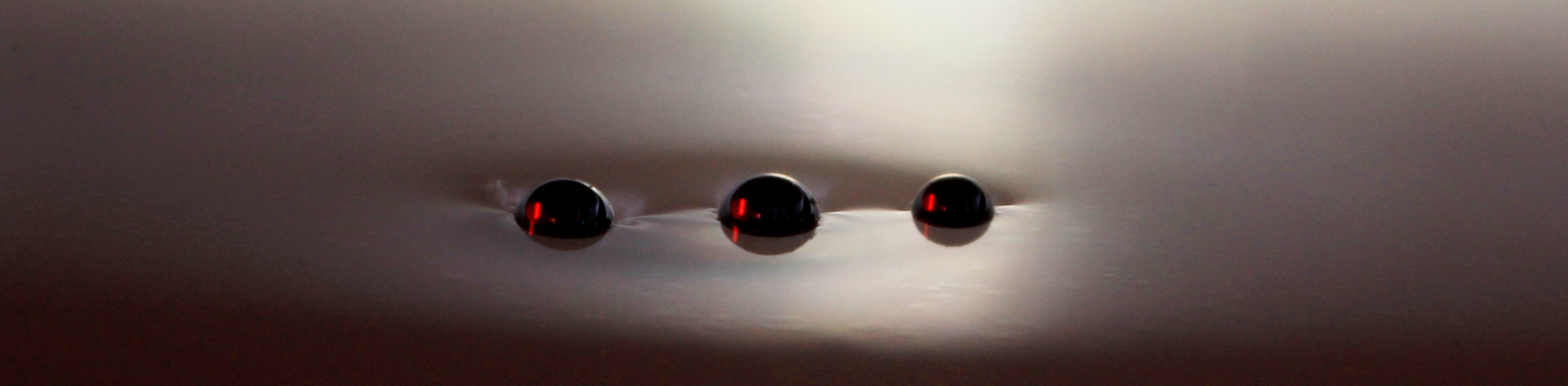}
\caption{Photograph of the magnetocapillary swimmer.
It is composed of two steel spheres of diameter $500~\mu$m and one of $397~\mu$m, partially immersed in water.
The meniscus around the spheres allows flotation and generates an attraction.
}
\label{photo}
\end{figure}

\begin{figure}
\includegraphics[width=8.6cm]{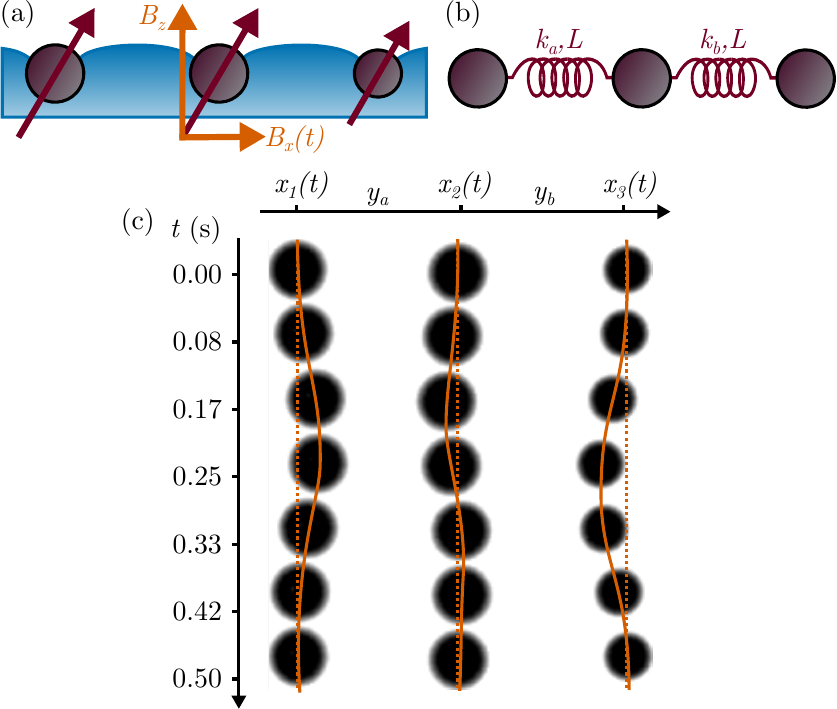}
\caption{(a) In the experiment, three beads experience a combination of magnetic dipole-dipole interactions and an attraction due to surface deformation.
(b) The model is composed of three aligned particles linked by two springs of rest length $L$ and spring constants $k_a$ and $k_b$.
(c) A spatio-temporal montage shows a non-reciprocal deformation sequence in the experiment.}
\label{schema}
\end{figure}

Three ferromagnetic steel spheres are placed at an air-water interface and exposed to magnetic induction fields using triaxis Helmholtz coils, as seen on Fig.~\ref{photo}.
These particles attract due to the deformation of the air-water interface they induce~\cite{vella2005}.
In the presence of a vertical induction field $B_z$, large enough to counter this attraction, the particles self-assemble~\cite{lagubeau2016,lumay2013,chinomona2015}.
This process is characterized by the magnetocapillary number $\mathcal{M}_c$, defined as the ratio between magnetic and capillary forces~\cite{lagubeau2016,chinomona2015}.
The self-assembly can reach two possible configurations : an equilateral one~\cite{grosjean2015} and a collinear one~\cite{chinomona2015}.
The latter is only stable with the addition of a horizontal field $B_x$ larger than a critical value $B_x^*$, which is a function of $\mathcal{M}_c$~\cite{chinomona2015}.
A sketch of this configuration is given in Fig.~\ref{schema}(a).
For a typical experimental value of $\mathcal{M}_c \approx 0.1$, the collinear state is stable for a relatively narrow range of parameters, as $B_x^*$ is close to the value for which contact occurs between the particles.

In the collinear state, particles are arranged similarly to the model shown in Fig.~\ref{schema}(b).
Indeed, the magnetocapillary interaction between two particles acts as a spring force for small displacements~\cite{lagubeau2016}.
Let us consider two oscillators with two different natural frequencies $f_a$ and $f_b$.
The interaction between the two outermost spheres is neglected.
This configuration allows to break time-reversal symmetry using a single excitation force, without needing independant forcings~\cite{golestanian2008, pickl2012, pande2015} or self-propelled components~\cite{babel2016}.
For this, particles of different diameters must be used.
Two spheres of diameter $D=500~\mathrm{\mu m}$ and one of $D=397~\mathrm{\mu m}$ are used throughout this letter, but different combinations of sizes have been tried with similar results.
We therefore have two magnetocapillary bonds: bond ``500-500" of elongation $y_a$ and spring constant $k_a$; and bond ``500-397" of elongation $y_b$ and spring constant $k_b$.
The corresponding natural frequencies are $f_a=1.810$~Hz and $f_b= 2.093$~Hz.
Rest lengths of the bonds are approximately equal, denoted $L$~\cite{vandewalle2013}.
In the experiments, we typically have $L \approx 2D$.
If each particle has a mass $m$ and a viscous damping coefficient $\mu$, we find $\mu/m = 14.6~\mathrm{s^{-1}}$ for $397~\mathrm{\mu m}$ beads and $\mu/m = 9.2~\mathrm{s^{-1}}$ for $500~\mathrm{\mu m}$ beads.
The quality factor $Q=\pi f m/\mu$ associated with each oscillator is $Q_a=1.09$ and $Q_b=1.89$.
Note that the quality factor must be non zero for at least one of the oscillators, so that a phase difference that is not a multiple of $\pi$ can appear around the resonance frequency.

When a time dependent horizontal field $B_x (t) = B_{x, 0} + \delta B \sin(\omega t)$ is added, with $\omega=2\pi f$, the interdistances change periodically, which can generate locomotion.
A spatio-temporal diagram illustrating the dynamics is given in Fig.~\ref{schema}(c).
Please note the phase shift in the elongation of oscillators $a$ and $b$.
Thanks to this phase shift, locomotion on the surface is possible.
Indeed, despite a reciprocal evolution of $B_x(t)$, the subsequent dynamics of the beads is non-reciprocal.

\begin{figure}
\includegraphics[width=8.6cm]{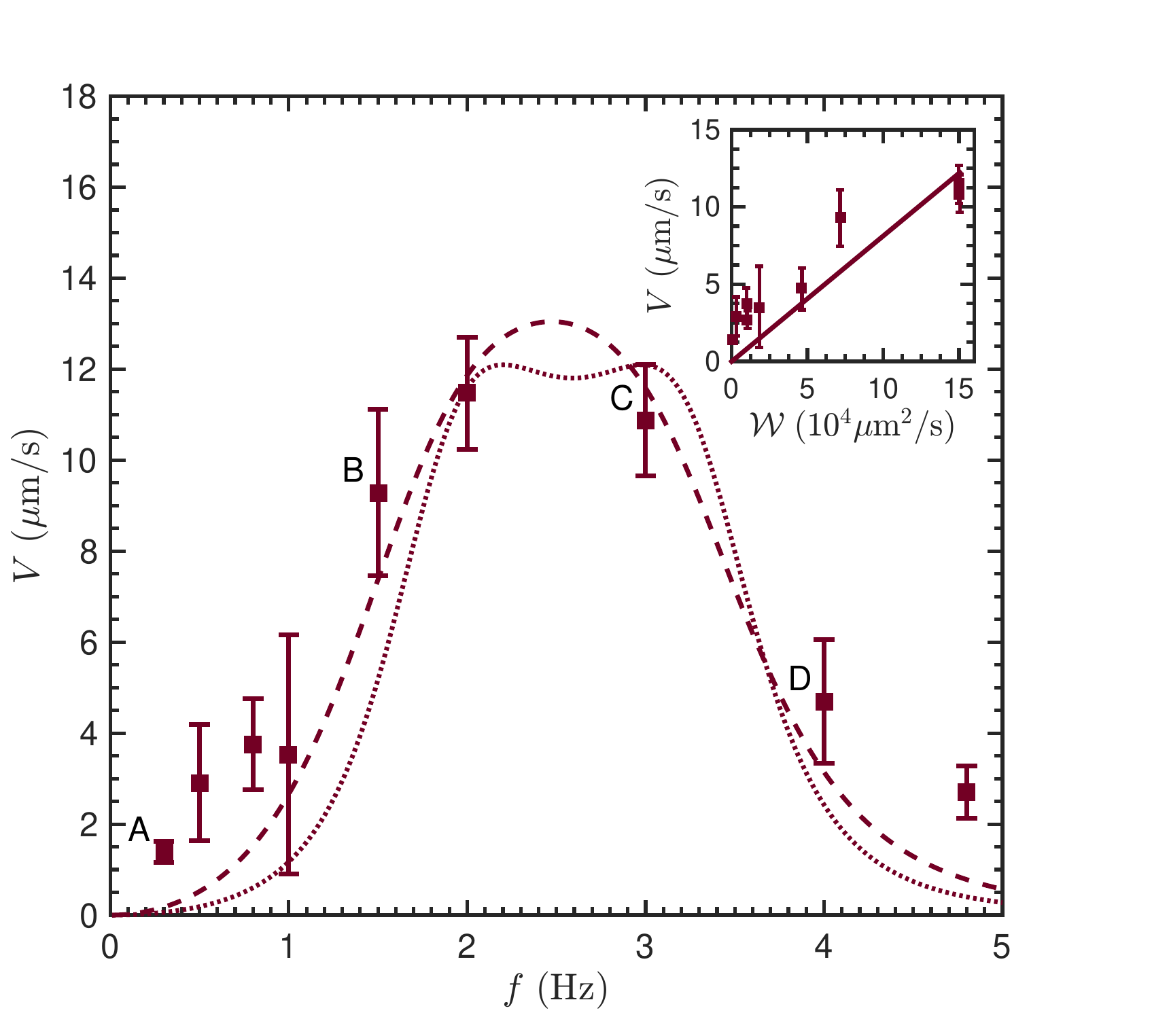}
\caption{Experimental swimming speed $V$ is plotted against the excitation frequency $f$.
Error bars represent the standard deviation on three experiments.
Dashed and dotted lines account for the model (Eq.~\ref{eq:wtheory}) with $397~\mathrm{\mu m}$ spheres and $500~\mathrm{\mu m}$ spheres respectively.
The linear relation between speed $V$ and efficiency $\mathcal{W}$ is illustrated in the inset.
}
\label{speed}
\end{figure}

Figure~\ref{speed} shows the swimming speed as a function of the excitation frequency $f$.
Each point is averaged over three independent experiments, for a total of 27 measurements. 
For each measurement, between 20 and 60 oscillation periods are recorded, depending on $f$.
We have $B_z = 4.5$~mT, $B_{x, 0} = 2.2$~mT and $\delta B \approx 0.5$~mT.
For low frequencies, the speed is almost equal to zero.
As the frequency $f$ approaches the natural frequency of the oscillators, the speed increases.
A maximum speed of around $12~\mathrm{\mu m/s}$ is typically reached between 2 and 3~Hz.
Above 3~Hz, the speed drops drastically.
Above 5~Hz, speed remains close to zero.

Note that if the amplitudes of the oscillating motions are too high, the swimmer can leave the region of stability of the collinear state, causing contact between the spheres or reaching the equilateral state~\cite{chinomona2015}.
This limits the possible values of amplitude $\delta B$, and, in turn, swimming speed.
It is possible, however, to further increase $\delta B$ by using a confinement potential to maintain the swimmer in the collinear state.
By placing the swimmer in a rectangular dish, with a concave meniscus perpendicular to the swimming direction, we can obtain such a confinement.
Higher speeds up to $76~\mathrm{\mu m/s}$ were obtained in this case.

\begin{figure}
\includegraphics[width=8.6cm]{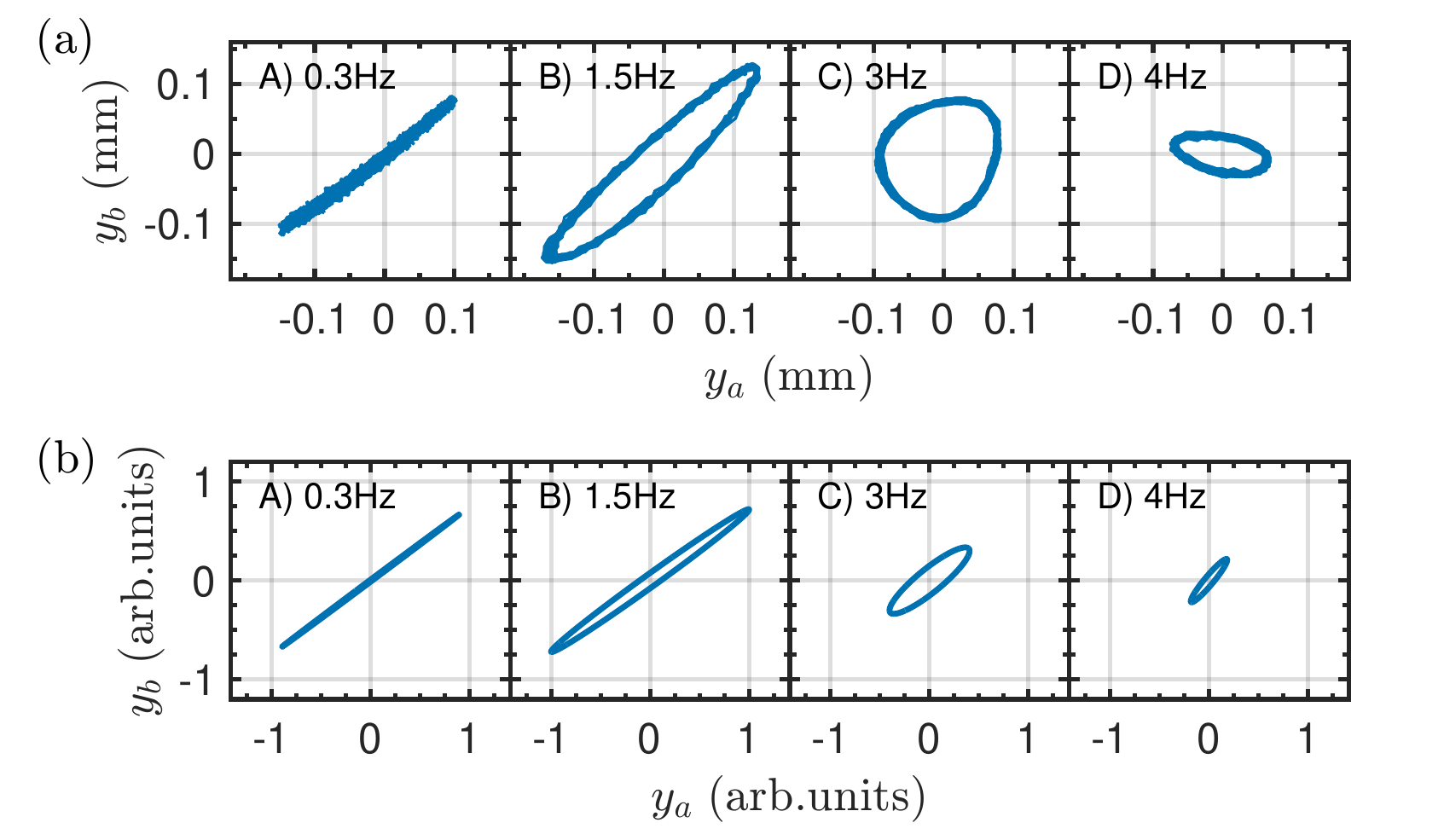}
\caption{(a) Experimental deformation cycles in the plane $(y_a,y_b)$, for four values of the excitation frequency.
(b) Deformation cycles predicted by the model.
Units are arbitrary, as the solutions depend linearly on the excitation amplitude.}
\label{cycles}
\end{figure}

In the kinematic model of Najafi and Golestanian~\cite{najafi2004}, the swimming speed is given by
\begin{equation}
V = \alpha A_{a}A_{b}\omega\sin\left(\phi_a-\phi_b\right) = \alpha\mathcal{W}
\label{eq:efficiency}
\end{equation}
where $A_a$ and $A_b$ are the amplitudes of elongation and $\phi_a$ and $\phi_b$ are their respective phase relative to the external field.
This means that the swimming direction is determined by the sign of $\phi_a - \phi_b$.
We can define the swimming efficiency $\mathcal{W}=A_a A_b \omega \sin\left(\phi_a-\phi_b\right)$.
The linear relation between $V$ and the experimentally measured value of $\mathcal{W}$ is plotted in the inset of Fig.~\ref{speed}.
The proportionality coefficient $\alpha$ given by a fit on the experimental data is $\alpha_{\mathrm{expt}} = 8.46\pm2.44\times10^{-5}~\mathrm{\mu m^{-1}}$.
According to~\cite{golestanian2008}, $\alpha$ can be expressed as a function of bead diameters and interdistances.
Using our experimental parameters, and with $L = 10^{-3}$~m, we find $\alpha_{\mathrm{th}} = 6.68\times10^{-5}~\mathrm{\mu m^{-1}}$, which lies in the 95\% confidence interval of the fit.
As the spheres are partially immersed, we can correct the Stokes force by a factor 0.86 as determined in~\cite{lagubeau2016}.
By adding this correction to the expression in~\cite{golestanian2008}, we find $\alpha_{\mathrm{th}} = 7.77\times10^{-5}~\mathrm{\mu m^{-1}}$.

Experimental deformation cycles, in the plane defined by the elongations $y_a$ and $y_b$, are depicted in Fig.~\ref{cycles}(a).
The four cycles correspond to four points in Fig.~\ref{speed}.
It can be shown that the area enclosed in the elliptical trajectory of the oscillators in the $(y_a,y_b)$ plane is given by $\pi\mathcal{W}/\omega$.
It is therefore proportional to the swimming speed.
Cycle A shows that the oscillators are in phase at frequencies close to zero.
This motion is reciprocal, such that the speed is also close to zero in Fig.~\ref{speed}.
The cycles open up progressively, between 0.5 and 3~Hz, resembling an ellipse as oscillators get out of phase.
This is correlated with the increase in speed observed in Fig.~\ref{speed}.
Around 3~Hz, oscillators are in quadrature and cycle C is approximately circular.
This is the optimal phase for swimming, but not necessarily where the maximum speed is reached, as seen in Fig.~\ref{speed}.
Indeed, oscillation amplitudes decrease with increasing frequency, which in turn decreases $\mathcal{W}$.
At higher frequencies, the oscillators are close to being in phase opposition, as seen on cycle D.
Furthermore, oscillation amplitudes decrease to zero, with the oscillation of bond 500-397 ($y_b$) decreasing faster than which of bond 500-500 ($y_a$).
This is correlated with a decrease in speed in Fig.~\ref{speed}.

Let us now investigate the model sketched in Fig.~\ref{schema}(b), which consists of three particles linked by two linear springs.
For the sake of simplicity, each particle has the same mass $m$ and experiences the same viscous damping $\mu$.
Oscillators also have the same natural length.
The Reynolds number in the experiment is defined as Re $=DAf/\nu$ where $\nu$ is the kinematic viscosity of water.
It is typically comprised between $10^{-3}$ and $10^{-1}$, meaning that the viscous dissipation dominates over inertia in the flow.
However, the inertia of the particles is not neglected in the model.
Indeed, the quality factor of each oscillator is close to 1, meaning that the oscillators are close to critical damping~\cite{lagubeau2016}.
The equations of motion are obtained via Newton's law
\begin{align}
&m \ddot{x}_1 + \mu \dot{x}_1 - k_a\left(x_2-x_1-L \right) = -F\sin\left(\omega t\right),\nonumber\\*
&m \ddot{x}_2 + \mu \dot{x}_2 + k_a\left(x_2-x_1-L \right) - k_b\left(x_3-x_2-L \right) = 0,\nonumber\\*
&m \ddot{x}_3 + \mu \dot{x}_3 + k_b\left(x_3-x_2-L \right) = F\sin\left(\omega t\right),
\label{eq:Newton}
\end{align}
where $F\sin\left(\omega t\right)$ is the external forcing at angular frequency $\omega$ and amplitude $F$.
This forcing is identical for each pair of beads.
As expected, the central particle is not submitted to any net forcing.
Please note that the sum of all internal and external forces is equal to zero.
We will now study the oscillators in terms of the elongations $y_a$ and $y_b$.
Defining $\underline{t} = \omega t$, $\Omega_a = k_a/m\omega$, $\Omega_b = k_b/m\omega$ and $\beta = \mu/2m\omega$, one has 
\begin{equation}
\begin{aligned}
\overline{\overline{y}}_a + 2\beta \overline{y}_a + 2\Omega^2_a y_a - \Omega^2_b y_b &= \sin\left(\underline{t}\right),\\*
\overline{\overline{y}}_b + 2\beta \overline{y}_b + 2\Omega^2_b y_b - \Omega_a^2 y_a &= \sin\left(\underline{t}\right),
\end{aligned}
\label{eq:dimensionlessNewton}
\end{equation}
where overlined symbols are derived by means of $\underline{t}$ and the elongations $y$ are expressed in $m\omega^2/F$ units.
The whole dynamics is described thanks to three dimensionless parameters: natural frequencies $\Omega_a$ and $\Omega_b$, and viscous damping $\beta$.
Those equations can be studied in Fourier space by considering the complex amplitudes of oscillation $\hat{y}_a = A_a\exp(-i\phi_a)$ and $\hat{y}_b = A_b\exp(-i\phi_b)$.
The solutions for both oscillators are
\begin{equation}
\hat{y}_{a,b} = \frac{3\Omega_{b,a}^2-1 - 2 i \beta}{\left( 2\Omega_a^2 -1 - 2 i \beta \right)	\left( 2\Omega_b^2 -1 - 2 i \beta \right)-\Omega_a^2\Omega_b^2	}.
\end{equation}
From this, we can find an expression for amplitude $A$ and phase $\phi$.

Figure~\ref{cycles}(b) shows cycles of deformation in the plane $(y_a, y_b)$ for typical parameters encountered in the experiments~\cite{lagubeau2016}.
Several features of the experimental cycles are recovered.
Indeed, low frequency cycles are similar.
Cycles gradually open up as frequency $f$ is increased.
The optimal phase difference is reached around 3~Hz.
However, amplitude decreases with $f$, such that speed is maximal between 2 and 3~Hz, as will be shown below.
At higher frequencies, the shape of the cycles are less accurately predicted, as can been seen on cycles C and D.
However, the model considers three identical spheres and neglects hydrodynamic couplings.
Indeed, around the resonance frequencies, there is typically a factor 10 between the hydrodynamic coupling and the restoring force.
Both the effect of size on viscous drag and the presence of hydrodynamic interactions could explain why phase difference is larger in the experiment, especially at higher frequencies.
It can be shown that the effect of the hydrodynamic coupling on the phase scales as $\omega$.
As observed experimentally, a non-reciprocal dynamic is observed despite the reciprocal evolution of the field $B_x(t)$.
The toy model rationalizes this observation: the distinct resonant frequencies $f_a$ and $f_b$ provide the spatial symmetry breaking required for the non-reciprocal deformation.

Let us quantify the velocity in the swimming regime as a function of the three dimensionless parameters.
In the Fourier formalism, Eq.~(\ref{eq:efficiency}) reads 
\begin{equation}
	\mathcal{W} = \text{Im}\left(\hat{y}_a^\dagger\hat{y}_b\right),
\end{equation}
which leads to the following dimensionless expression
\begin{equation}
\mathcal{W} = \frac{6\beta\left(\Omega_b^2-\Omega_a^2\right)}{\Big\vert\Omega_a^2\Omega_b^2-\left(\left(2\Omega_a^2-1\right)+2i\beta\right)\left(\left(2\Omega_b^2-1\right)+2i\beta\right)\Big\vert^2},
\label{eq:wtheory}
\end{equation}
with $i$ being the complex unit. Note that this equation is anti-symmetrical with respect to oscillators $a$ and $b$ and, consequently, leads to zero if both oscillators are identical.
Indeed, in this case, the oscillations would be in phase.
Note also that with no damping, \emph{i.e.} with $\beta=0$, the velocity is also zero, as expected.
This equation also indicates the direction of motion, given by the sign of $\Omega_a^2-\Omega_b^2$.

\begin{figure}
\includegraphics[width=8.6cm]{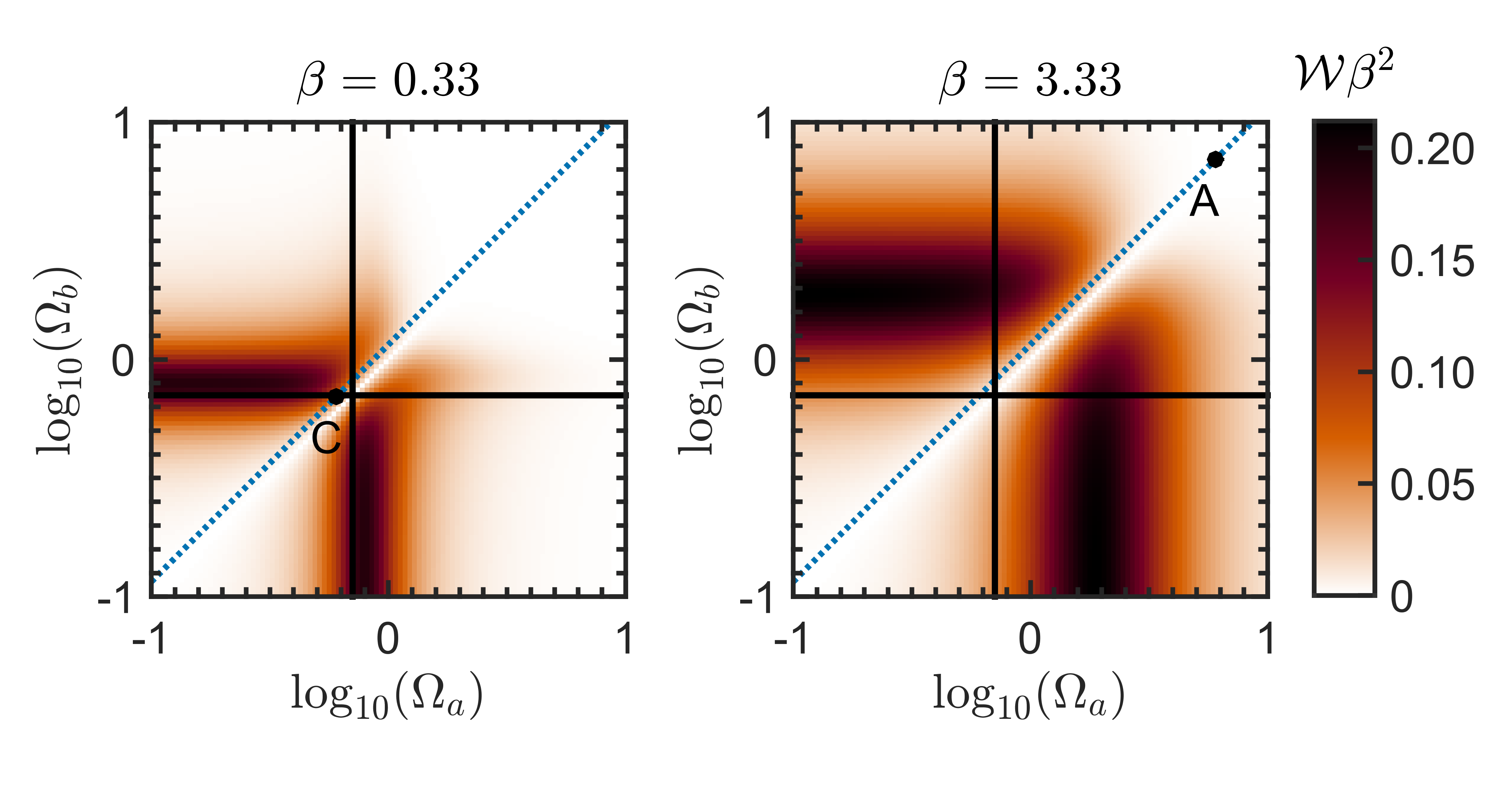}
\caption{Theoretical prediction for efficiency $\mathcal{W}$ as a function of the dimensionless excitation frequencies.
Two values of damping $\beta$ are shown.
The continuous lines represent the undamped resonance frequencies, while the dotted line represents a sweep of frequency $f$ in the experiment.
It is close, but distinct from the identity line.
Two points corresponding to the cycles A and C of Fig.~\ref{cycles} are placed on each plot, as $\beta$ is a function of $f$.}
\label{maps}
\end{figure}

Figure~\ref{maps} shows the effect of $\Omega_a$ and $\Omega_b$ on the efficiency $\mathcal{W}$ for two values of $\beta$.
When $\beta$ is close to zero, two sharp lines of high efficiency $\mathcal{W}$ are observed at $\Omega_{a,b} \approx 1/\sqrt{2} = \Omega_{a,b}^{\mathrm{res}}$, due to the resonance of each oscillator.
Speed is zero along the diagonal line $\Omega_a=\Omega_b$, corresponding to identical oscillators.
As $\beta$ increases, the maximum efficiency decreases roughly like $\beta^{-2}$.
The optimal region widens and shifts towards higher values of $\Omega$.
This comes from the decrease in the oscillation amplitudes $\hat{y}_{a,b}$.
Two points that correspond to two cycles of figure~\ref{cycles} are shown.
They are in a diagonal line that is close, but distinct to the identity line.
Each point is associated with a different value of $\beta$, as $\beta$ is a function of the excitation frequency.
As can been seen from their position in the graphs, cycles C and A correspond to an efficient and an inefficient swimmer, respectively.

The theoretical expression of $\mathcal{W}$ is compared to the experiment in Fig.~\ref{speed}.
Experimental parameters are injected in Eq.~(\ref{eq:wtheory}), leaving no fitting parameter aside from a vertical scaling.
One observes that the model reproduces correctly the experimental observations, despite the approximations made in order to reach an analytical expression for $\mathcal{W}$.
As the model considers identical spheres, the predictions for both $397~\mu$m and $500~\mu$m spheres are shown.

The experiment was conducted with submillimeter-sized particles, which are on the larger end of the spectrum of low Reynolds swimming.
One could wonder how efficiency $\mathcal{W}$ would be affected by a downscaling.
Decreasing the size of the spheres and the distances between them changes the values of $\beta$ as well as the resonance frequencies of the magnetocapillary bonds~\cite{lagubeau2016}.
Let us assume that all length scales decrease with $D$ and that forcing frequency remains close to the resonant frequencies.
Thanks to dimensional analysis, one finds that the velocity of a magnetocapillary swimmer scales as $V \sim D^{-1}$.
This suggests that a downscaled version of the swimmer would be able to propel itself effectively.
However, particles smaller than $3.4~\mu$m would experience a capillary force weaker than thermal agitation~\cite{lagubeau2016}.
The deformation of the liquid surface around the particles could be enhanced to lower this bound, for instance through geometrical constraints, by using denser, more hydrophobic particles or by applying a vertical force other than gravity on the particles.

In summary, we realized experimentally the Golestanian-Najafi swimmer using ferromagnetic spheres linked by magnetocapillary bonds.
Thanks to the dynamics of the bonds, the deformation is non-reciprocal and therefore induces a motion of the swimmer along the surface.
A linear toy model has been developed that reproduces the speed profile, as well as the general behavior of the deformation cycles.
More specifically, we obtained the expression of the swimmer velocity as a function of the fluid parameters and the resonance frequencies of the magnetocapillary bonds.
The theoretical expression of the velocity suggests that this swimmer could be efficient at even smaller scale.

As linear springs are used to model the magnetocapillary bonds, the model remains general.
The same approach could be applied to various systems, using other restoring forces.
For instance, the same particles could be physically linked by an elastic material.
This would allow experiments in the bulk and could prove more robust than a self-assembly.
The model uses two different spring constants to generate the breaking of symmetry under a uniform forcing, which could also be obtained with materials of different elasticity.
Furthermore, it can be shown that a similar spontaneous phase shift can be obtained by using different viscous dampings and/or particle masses.
This approach could serve to generate various assemblies more suited to technological applications requiring microswimmers, micromanipulators or micropumps.

\acknowledgments
This work was financially supported by the University of Li\`ege (Grant FSRC 11/36). GG thanks FRIA for financial support.

\bibliography{biblio}{}
\bibliographystyle{apsrev4-1}

\end{document}